\documentclass[conference]{IEEEtran}

\IEEEoverridecommandlockouts
\usepackage[disable]{todonotes}

\begin{document}
\title{Speculative Leakage in ARM Cortex-A53
}

\author{
 \IEEEauthorblockN{Hamed Nemati\IEEEauthorrefmark{1},
                   Roberto Guanciale\IEEEauthorrefmark{2},
                   Pablo Buiras\IEEEauthorrefmark{2},
                   Andreas Lindner\IEEEauthorrefmark{2}}

 \IEEEauthorblockA{\IEEEauthorrefmark{1}Helmholtz Center for Information Security (CISPA)
    \\ hnnemati@cispa.saarland}
                   
 \IEEEauthorblockA{\IEEEauthorrefmark{2}KTH Royal Institute of Technology
    \\\{lindnera, buiras, robertog\}@kth.se}

}

\maketitle

\newcommand{\vulnerabilityname}{SiSCloak}

\begin{abstract}
The recent Spectre attacks have demonstrated that modern microarchitectural optimizations can make
software insecure. These attacks use features like
pipelining, out-of-order and speculation to extract
information about the memory contents of a process via side-channels.

In this paper we demonstrate that Cortex-A53 is affected by speculative leakage
even if the microarchitecture does not support out-of-order execution. We named this new class of vulnerabilities \vulnerabilityname\footnote{SIngle SpeCulative LOad AttacK}.
\end{abstract}

\section{Introduction}
Both Spectre and Meltdown \cite{DBLP:conf/sp/KocherHFGGHHLM019,lipp2018meltdown} have provided evidence of the fundamental insecurity of current computer microarchitecture. The use of instruction level parallelism, out-of-order, and speculative execution has produced processor designs with side channels that can leak sensitive information about the memory contents of programs.
These attacks use cache side channels that can be measured via several methods, as Prime+Probe~\cite{liu2015last} and Flush+Reload~\cite{yarom2014flush}.

When the first Spectre attack was published, some microarchitectures
(e.g., Cortex-A53\todo{P:unify notation: Cortex A53 or Cortex-A53 (with
  hyphen)}) were claimed immune to these types of attack because they allow speculative fetching but not speculative
execution or out-of-order execution.
The informal argument was that mispredictions cannot cause buffer overreads or
leave any footprint on the cache in the absence of
speculative loads.

In this paper we show that these architectures can be affected by a new class of vulnerabilities, dubbed \vulnerabilityname, and
we demonstrate them on ARM Cortex-A53.

\section{Spectre-PHT}\label{sec:spectre:v1}
The original Spectre vulnerability, dubbed Spectre-PHT~\cite{DBLP:conf/sp/KocherHFGGHHLM019}, exploits the prediction
mechanism for the outcome of conditional branches.
Modern CPUs use \emph{Pattern History Tables} (PHT) to record
patterns of past executions of conditional branches, i.e.,
whether the \emph{true} or the \emph{false} branch was executed, and
then use it to predict the outcome of that branch.
By poisoning the PHT to execute one direction (say the
\emph{true} branch), an attacker can fool the
prediction mechanism into
executing the \emph{true} branch, even when the actual outcome of the branch is ultimately \emph{false}.
The following program illustrates
information leaks via Spectre-PHT: \\
\begin{verbatim}
R1 = LD[#A-size]
if (R0 < R1)
  R2 = LD[#A+R0]
  R3 = LD[#B+R2]
\end{verbatim}
In this case the victim owns arrays \verb|A| and \verb|B| that
start at address \verb|#A| and \verb|#B| respectively.
The size of the first array is stored at address \verb|#A-size|. We assume that the two arrays do not
contain confidential data, that every element of \verb|A| is a valid index into the array \verb|B|, and that
the attacker controls the value of register \verb|R0|.

This program is considered secure at the ISA level as it ensures that \verb|R0| always lies within the bounds of \verb|#A| and
the three memory accesses are only dependant of public information: the location \verb|#A-size|, the locations of  \verb|#A| and \verb|#B|,
the attacker input \verb|R0|, and \verb|#A[R0]|.

However, if the microarchitecture supports speculative execution, an attacker can fool the prediction mechanism by first supplying values of
\verb|R0| that execute the \emph{true} branch, and then a value that
exceeds the size of \verb|#A|. This causes the CPU to perform an
out-of-bounds memory access in \verb|LD[#A+R0]| that reads sensitive data, which is later
used as index for a second memory read \verb|LD[#B+R2]|. The latter access can leak the sensitive data
by leaving a trace on the cache.

\section{Invulnerability of ARM Cortex-A53}\label{sec:spectre:v1}
Some microarchitectures
(e.g., Cortex-A53) were claimed immune to Spectre attacks because they allow speculative fetching but prevent speculative
execution of the fetched instructions. The variation of Spectre presented above is not successful on these architectures because
the CPU may mispredict the branch output, fetch and decode the instructions in the wrong branch, but it cannot execute the two memory reads.
This prevents the speculative leakage.

During our experiments we discovered that the above claim is not totally accurate. For instance, Cortex-A53 is actually capable of
executing instructions in speculation, but it does not allow to use the result of a speculated instruction for subsequent operations, probably
due to the absence of register renaming and the short CPU pipeline.
For instance the CPU may speculatively request the memory subsystem to load from \verb|#A+R0|,
but the result of the memory operation cannot be used in speculation, hence \verb|LD[#B+R2]| is stalled until the branch condition
is resolved.
For this reason, the load from \verb|#A+R0|, whose address is not confidential, can affect the cache, while
the load from \verb|LD[#B+R2]|, whose address may contain confidential data, cannot.
This makes the original version of Spectre ineffective on ARM Cortex-A53.

\section{\vulnerabilityname}
The discussion above does not answer a more general question: do Cortex-A53
cores prevent all leakage due to speculation?
Here we show that the answer is no and we provide two similar counterexamples.
The first counterexample is the following program:
\begin{verbatim}
R1 = LD[#A-size]
if (R0 < R1)
  R2 = LD[#A+R0]
  if (R2 & 0x80000000)
    R3 = LD[#B+R2]
\end{verbatim}
In this case we assume that array \verb|B|
does not contain confidential data, that every element of \verb|A| is a valid index into the array \verb|B|, and that
the attacker controls the value of register \verb|R0|.
We also assume that the highest bit of each element of \verb|A| identifies the classification of the element itself, i.e.
the element is public only if the highest bit is set.
This program is considered secure at the ISA level as it ensures that \verb|R0| always lies within the bounds of \verb|#A| and
that no memory access depends of non-public information.
However, we discovered that on Cortex-A53 this program is not secure. In fact, the CPU may mispredict condition \verb|R2 & 0x80000000|
which leads to consider a confidential element is public. In this case, the CPU may speculatively access \verb|#B+R2|, making the confidential
data in \verb|R2| affect the cache.

The second counterexample (which was also previously presented but not experimented in~\cite{guarnieri2020hardwaresoftware}) is a variation of Spectre PHT:
\begin{verbatim}
R1 = LD[#A-size]
R2 = LD[#A+R0]
if (R0 < R1)
  R3 = LD[#B+R2]
\end{verbatim}
With respect Spectre PHT, the access \verb|R2 = LD[#A+R0]| has been anticipated by the programmer
or the compiler.
In this case, a Cortex-A53 CPU may mispredict condition \verb|R0 < R1|
and speculatively access \verb|#B+R2|, which may contain data that has been read out-of-bound.

\section{Experimentation}
We conducted our experiments on Raspberry Pi 3, which is a widely available ARMv8 embedded system. The platform's CPU is a Cortex-A53, 
and according to the reference manual it is an 8-stage pipelined processor with a 2-way superscalar and in-order execution pipeline. 
The CPU also implements branch prediction and a \textit{performance monitor counter} (PMC) which we used for timing analysis.

Our experimentation platform runs as bare-metal code, there are no background processes or interrupts. In fact, our goal is to show that
these speculative leakages exist and we leave their practical exploitation as future work.
Similar to the original Spectre PHT attack we mistrain the branch predictor by executing the victim code in a loop with valid inputs, thereby training
the branch predictor to expect that conditional expression will be true. Next we execute the victim code with malicious inputs. While according to 
the processor reference manual speculative execution of the code residing in true branch should not happen with malicious input, we observed different 
effects at the microarchitectural level. 

In order to collect evidence that speculative execution of code after the conditional expression happens, we initially used ARM TrustZone and we exploited privileged debug instructions to directly inspect the cache state.
Later we simulated a real attack,  were we used timing analysis to leak the secret. This is done using Flush+Reload attack, where the attacker invalidates  shared cache lines, causes the execution of the victim code, and then measures the timing of a later access to the shared lines. If the access time to the shared lines incurs shorter latency, the attacker can infer which lines have been accessed by the victim.
To access timing information we used the cycle counter of Cortex-A53 PMC, which allows us to accurately  measure latency of memory accesses.

\section{Concluding Remarks}
We show two new speculative vulnerabilities that affect the ARM Cortex-A53 processor. This invalidates common beliefs regarding the immunity of this processor against Spectre-like attacks. Our experiments show that Cortex-A53 speculatively executes instructions located after a conditional expression which needs to load its operands from the memory. However, the length of this speculative execution is limited to at most two instructions; if one of the two instructions is a load from a cacheable memory address then the corresponding cache-line is speculatively filled with values from the memory.

Recently ARM published a similar vulnerability, which has been called ``straight-line speculation''~\cite{straight}.
Straight-line speculation involves the processor to speculatively execute the next instructions past an unconditional change in control flow. The report shows that existing ARM processors are affected by this vulnerability for
unconditional indirect branches and function returns, but are not affected for unconditional direct branches. We believe that both \vulnerabilityname\ and straight-line speculation are caused by the ability of ARM Cortex-A53 to speculatively issue memory requests when the control flow can not be statically determined.

\section*{Responsible Disclosure}
\vulnerabilityname\  was responsibly disclosed to ARM by the authors on June 2020. As part of the disclosure process, ARM confirmed that the Cortex-A53 is vulnerable to attacks based on single speculative memory loads, as described in this report.
\bibliographystyle{splncs04}
\bibliography{biblio}

\end{document}